# Ensemble Chinese End-to-End Spoken Language Understanding for Abnormal Event Detection from Audio Stream

H., WEI*

Department of ECE, University of Texas at Dallas, haoran.wei@utdallas.edu

Seattle AI lab, Kuaishou Technology

F. , TAO

Seattle AI lab, Kuaishou Technology, feitao@kuaishou.com

R. , SU

Department of Statistics and Probability, Michigan State University

Seattle AI lab, Kuaishou Technology, runzesu@kuaishou.com

S. , YANG

AI platform, Kuaishou Technology, senyang@kuaishou.com

J. , LIU

Seattle AI lab, Kuaishou Technology, jiliu@kuaishou.com

Conventional spoken language understanding (SLU) consist of two stages, the first stage maps speech to text by automatic speech recognition (ASR), and the second stage maps text to intent by natural language understanding (NLU). End-to-end SLU maps speech directly to intent through a single deep learning model. Previous end-to-end SLU models are primarily used for English environment due to lacking large scale SLU dataset in Chinese, and use only one ASR model to extract features from speech. With the help of Kuaishou technology, a large scale SLU dataset in Chinese is collected to detect abnormal event in their live audio stream. Based on this dataset, this paper proposed a ensemble end-to-end SLU model used for Chinese environment. This ensemble SLU models extracted hierarchies features using multiple pre-trained ASR models, leading to better representation of phoneme level and word level information. This work also showed a feasible way to build a end-to-end SLU system without transcribing audios. The proposed approached achieve 9.7\% increase of accuracy compared to previous end-to-end SLU model.

CCS CONCEPTS • Computing methodologies • Artificial intelligence • Natural language processing

**Additional Keywords and Phrases:** End to end spoken language understanding, abnormal event detection from audio stream, ensemble Chinese spoken language understanding

---

* Correspondence: haoran.wei@utdallas.edu

# 1 INTRODUCTION

With the rapid growing of live-streaming applications, like Facebook Live and Kwai, semantics information from these applications need to be utilized to detect abnormal events. Spoken language understanding (SLU) system is used to infer the intent of a speech and to detect the abnormal events.

Conventional SLU system typically consist of two stages, automatic speech recognition (ASR) stage converts speech to text, then the natural language understanding (NLU) stage extract the intent from the recognised text [1, 2, 3]. The limitations of conventional SLU system include: 1)ASR and NLU modules are optimized separately, then NLU part suffers from ASR errors; 2) Information helpful for inferring the intent like prosody is not present in text transcript; 3) ASR and NLP models including search algorithms, language models, finite state transducers are very large and complex.

To address the limitations mentioned above, End-to-end SLU was proposed, End-to-end SLU maps speech directly to intent through a single deep learning model [4, 5, 6, 7, 8, 9, 10, 11, 12].

Due to lacking large scale SLU dataset in Chinese, previous end-to-end SLU models are primarily used in English and French environment [13]. With the help of Kuaishou technology, a large scale SLU dataset in Chinese is collected to detect abnormal event from their audio live-streaming. Based on this dataset, this paper proposed a ensemble end-to-end SLU model used in Chinese environment.

Ensemble of multi-modalities have been proved to generate better results compared to using only one modality in many tasks [14, 15, 16, 17, 18]. This ensemble end-to-end SLU models are able to extract hierarchical features using multiple pre-trained ASR, leading to better representation of phoneme level and word level information from speech.

The contribution of this paper includes:

1. A large scale Chinese SLU dataset is collected by Kuaishou technology. Based on this dataset, this paper proposed a Chinese end to end SLU system to detect abnormal event from audio live-streaming application.

2. This is the first paper to develop spoken language understanding (SLU) task without transcribing the audios. All previous work requires speech transcription.

3. The ASR part of this paper is embedded with ASR ensemble output. This ensemble ASR embedding provide a better representation of the speech.

4. Unlike previous SLU tasks, the ensemble end-to-end SLU model use super long duration speech of 15 seconds. In this scenario, attention layer can not perform well. A max-pooling layer is attached to replace the attention layer, which achieve better performance.

The rest of the paper is organized as follows: a description of related works and related datasets are provided in Section 2. The Chinese large scale SLU dataset for abnormal event detection and the proposed ensemble end-to-end SLU approach are described in Section 3. Experimental setup and results are presented in Section 4. Finally, the conclusion and future work are summarized in Section 5.

# 2 RELATED WORKS

The conventional SLU architectures and end-to-end SLU architectures will be introduced in Section 2.1 and 2.2 respectively. The related datasets are described in Section 2.3.

## 2.1 Conventional SLU

Conventional SLU system typically consist of ASR part [19, 20, 21, 22] and NLU part [23, 24, 25]. Figure 1 shows the diagram of a conventional SLU system. ASR part aims to get word sequence output from speech input. There are two widely used end-to-end ASR architectures. The hybrid connectionist temporal classification (CTC)/attention end-to-end ASR [26], which utilizes the advantages of both CTC and attention based method. The other is transformer based ASR architecture, which learns sequential information via a self-attention mechanism instead of the recurrent connection [27].

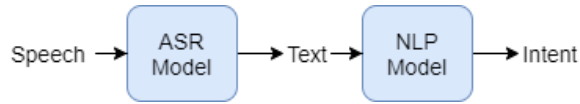

Figure 1: Diagram of the conventional SLU system.

To utilize these widely used ASR architectures by PyTorch, ASR toolkit ESPnet [28, 29, 30] are proposed. Deep learning engine part of ESPnet uses chainer [31] and PyTorch, and other parts also follows Kaldi [32] style data processing, feature extraction/format, and recipes.

Given the word sequence output from ASR part, the NLU part is utilized to predict the intent from the given word sequence. Paper [5] found two layer bi-LSTM (long short-term memory) encoder performs better in the conventional SLU system.

For the conventional SLU system, ASR and NLU parts are trained independently. During the training phase, NLU part is trained using clean transcription. During evaluation phase, the output of ASR is piped into the NLU part.

## 2.2 End-to-end SLU

Recently, end-to-end SLU has been utilized to maps speech directly to intent through a single deep learning model. Figure 2 shows the diagram of a end-to-end SLU system. For researches in this area, paper [5] proposed an end-to-end SLU without pre-trained ASR. Paper [11] also trained from scratch, further performance improvement is achieved by utilizing data augmentation. End-to-end SLU approaches without pre-trained ASR can attain fairly good results, but still not able to beat approaches with pre-trained ASR.

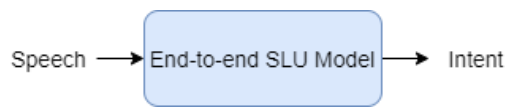

Figure 2: Diagram of the end-to-end SLU system.

Paper [8] uses pre-trained ASR to get the graphemes, then the softmax probabilities over graphemes output are fed directly into intent classifier. Similar to [8], paper [6] remove the restriction of the softmax bottleneck and uses alternative training targets of phonemes and words. Paper [6] also introduced a new SLU dataset, called Fluent Speech Commands dataset.

Other techniques have also been utilized to improve the performance of end-to-end SLU system. Paper [7] use speech synthesis to generate a large synthetic training data, but the possible utterance content should be already known before the speech synthesis. Paper [12] proposed a approach for jointly predicting the spoken words and intent labels. For utilizing this approach, the spoken words should already been known. Paper [9] improved the end-to-end SLU by combining self-attention acoustic classifier and text classifier, but the combined classifier is not an end-to-end approach.

### 2.3 Related Dataset

The most widely used dataset of end-to-end SLU is Fluent Speech Commands dataset [6]. Fluent Speech Commands dataset is composed of 16000 Hz single-channel .wav audio in English. This dataset contains 30043 audios with 77 speakers. Each audio is labeled with "action", "object" and "location" slots. There are 248 distinct sentences in this dataset, each sentence is spoken by multiple speakers in both training and testing set.

Another dataset used for end-to-end SLU is much smaller than Fluent Speech Commands dataset, it is called Snip SLU dataset [33]. It only contain 1660 audios, each corresponding to a different sentence, so the audios for testing has never heard before.

French MEDIA corpus [13, 34] are recorded in French environment. It has concepts of reference, relative time, locations and prices that are expressed by short semantically ambiguous words, which can also been used for SLU tasks. There are 300 dialogs in this dataset, and there are 15 utterance in each dialog.

All of the datasets mentioned above are recorded in English or French environment. Audio streams from these datasets are already segmented, and usually with duration of a few seconds.

## 3 PROPOSED DATASET AND APPROACHES

In this section, the Chinese large scale SLU dataset for abnormal event detection is introduced in Section 3.1. The proposed approach of inferring the intent from this dataset is described in Section 3.2.

### 3.1 Chinese Large Scale SLU Dataset

Because of lacking large scale SLU dataset in Chinese, the previous end-to-end SLU approaches are all used for English environment. With the help of Kuaishou technology, a large scale SLU dataset in Chinese is collected to detect abnormal event from their audio live-streaming.

This dataset is collected in a realistic scenario, each audio is collected from a live-streaming host, and the host will talk whatever they want in the live-streaming platform. The content or transcript of this dataset is unknown in this scenario. Some of these content are inappropriate, and are treated as abnormal event. For example, some live-streaming host may says "please go to our store by this link, do not use the kwai's e-commerce link" in Chinese, which is not allowed by the platform. Variety background noise will appear in this dataset, and the noise type is also unknown. Some hosts have strong accent, making this dataset extremely difficult to recognize by ASR.

The target intent of the abnormal event detection dataset have 2 classes consisting of positive and negative intents, the positive intent corresponding to abnormal event, which is prohibited by the kuaishou platform. While the negative intent corresponding to normal event. There are in total 52446 audios in this dataset, corresponding to 22578 positive audios and 29868 negative audios. Each audios of this dataset have duration of 15 seconds and have been converted to 16000 Hz single-channel .wav file.

Abnormal event detect from this dataset is a very challenging tasks. Unlike previous SLU dataset, which have short duration speech of 2 to 5 seconds and already been segmented, the speech in abnormal event detect dataset last for 15 seconds. In this event detect dataset, useful information relevant to the intent only takes a small portion, and lots of irrelevant information are mixed with the target intention.

### 3.2 Proposed Ensemble End-to-End SLU Approach

This paper proposed an ensemble end-to-end SLU approach to detect abnormal event from Chinese large scale SLU dataset. Compare with other end-to-end SLU approaches, this approach utilize multiple pre-trained ASR models and provide a better representation of input speech. As the transcriptions of this dataset were not available, this paper develop SLU task without transcribing the audios. Figure 3 shows the diagram of this ensemble end-to-end SLU system.

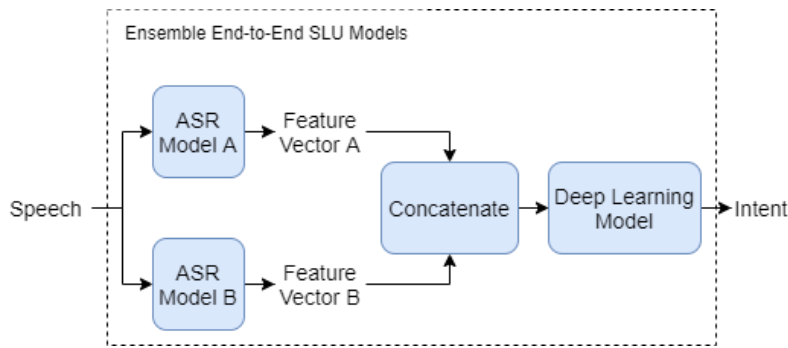

Figure 3: Diagram of the ensemble end-to-end SLU system.

The ASR model A used in this approach is an pre-trained hybrid CTC/attention end-to-end ASR. Feature vector A with 1024 dimensions are extracted from this ASR model A. The ASR model B used in this approach is an pre-trained transformer based ASR. Feature vector B with 256 dimensions are extracted from this ASR model B. Both of the ASR model are trained with ESPnet [28, 29, 30] utilizing all available Chinese corpora from OpenSLR. Data augmentation methods of reverberate and additive noises are also utilized to help the ASR training.

Feature vectors A and B from that two ASR models are concatenated, then fed into a deep learning model. This deep learning model consist of a 2 layers bi-directional LSTM, a liner layer and a max-pooling layer. The bi-directional LSTM have 256 hidden units in each layer. The linear layer map the output from each time step of LSTM to 2 intent classes. Output from each time step of linear layer are fed into a max-pooling operation alone the time axis. For speech input with very long duration, using max-pooling operation alone time axis is proved to be more efficient than using output from last time step only. Output of the max-pooling layer is the score for each intent classes.

### 4 EXPERIMENT AND RESULTS

To overcome the limitations associated with conventional SLU approach, end-to-end SLU approaches are proposed in recent study. Two end-to-end SLU approaches with different final layer are introduce first. Then the end-to-end SLU approach with better performance is compare with the ensemble end-to-end SLU approach.The experiment setup are illustrated in Section 4.1, then the experiment performance for each approaches are described in Section 4.2.

### 4.1 Experiment Setup

The end-to-end SLU approach utilize pre-trained hybrid CTC/attention end-to-end ASR to get feature vectors. Then this feature vector are fed into a deep learning model. This deep learning model consist of a 2 layers bi-directional LSTM, a liner layer and a final layer to deal with the output from different time steps. One approach use attention layer to fusion the output from each time steps, the other approach use max-pooling layer along the time axis to get the final output. Figure 4(a) shows the deep learning model with attention layer, and Figure 4(b) shows the deep learning model with max-pooling layer. Ensemble end-to-end SLU approach also use the same deep learning model as
Figure 4(b).

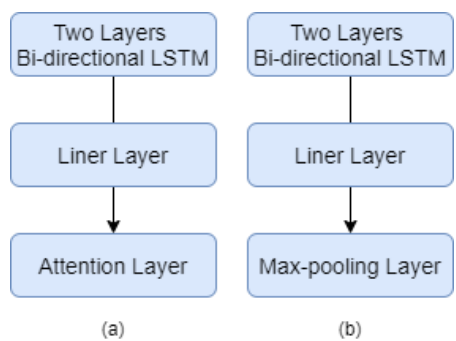

Figure 4: Deep learning model with different final layer, (a) deep learning model with attention layer, (b) deep learning model with max-pooling layer.

The pre-processing operations for all approaches are conducted in the Kaldi manner. More specifically, four files including spk2utt file, utt2spk file, wac.scp file and text file need to be prepared before the pre-processing. Based on the data augmentation and feature setting, speech are converted to .ark file and .scp file. Extracted features are stored in .ark file, and .scp file. provides location of extracted features in the ark file.

In this paper, reverberate and additive noises are utilized as the data augmentation methods to help the ASR training. The extracted features used in this paper is Filter Banks (FBank), which is proved to be more efficient than Mel-Frequency Cepstral Coefficients (MFCCs) features in recent deep learning based ASR tasks. To accelerate the data pre-processing operation, positive and negative audios are divided into 32 .ark files respectively. Audios from the front 30 .ark files are used for training, the rest audios from the last 2 .ark files are used for testing.

### 4.2 Results for Experiments

Comparison between end-to-end SLU approach with attention layer and end-to-end SLU approach with max-pooling layer is shown in Table 1. Then the performance of end-to-end SLU approach and ensemble end-to-end SLU approach are provided in Table 2.

As the abnormal event detection of these SLU models have 2 classes output, Recall, Precision and F1 score can be utilized to measure their performance.

In a two classes classification problem, true positives (TP) indicating correctly classified positive event, false positives (FP) indicating negative event misclassified as positive event, false negatives (FN) indicating positive event misclassified as negative event, and finally true negatives (TN) indicating correctly classified actions of negative event. Based on the numbers of TP, FP, FN, and TN, Recall, Precision and F1 score can be calculated as follows:

$$\text{Recall} = \frac{N_{TP}}{N_{TP}+N_{FP}} \quad (1)$$

$$\text{Precision} = \frac{N_{TP}}{N_{TP}+N_{FN}} \quad (2)$$

$$F1 = \frac{2 * \text{Recall} * \text{Precision}}{\text{Recall} + \text{Precision}} \quad (3)$$

In Table 1, end-to-end SLU with max-pooling layer get higher F1 score. This indicates max-pooling layer have better performance than attention layer for input sequence with super long time steps. For speech input with 15 seconds duration, it is difficult for attention layer to extract information from such long sequence.

Table 1: Performance of end-to-end SLU approaches with different final layer

| Model | Recall | Precision | F1 |
|---|---|---|---|
| End-to-end SLU with attention layer | 60.43% | 77.75% | 68.00% |
| End-to-end SLU with max-pooling | 65.45% | 88.19% | 75.14% |

As indicated in Table 1, end-to-end SLU approaches with max-pooling layer outperform the end-to-end SLU approaches with attention layer. Table 2 compared end-to-end SLU approach with ensemble end-to-end SLU approach, and both of the approaches use max-pooling layer. Table 2 shows that ensemble end-to-end SLU approach achieve the highest F1 score. The ensemble framework made it feasible to build end-to-end SLU system without transcribing the noisy audios.

Table 2: Performance of end-to-end SLU approach and ensemble end-to-end SLU approach

| Model | Recall | Precision | F1 |
|---|---|---|---|
| End-to-end SLU with max-pooling | 65.45% | 88.19% | 75.14% |
| Ensemble end-to-end SLU with max-pooling | 68.68% | 89.52% | 77.73% |

## 5 CONCLUSIONS

Conventional SLU approach have separate ASR stage and NLU stage. The limitations of this approach include large and complex model, discarding useful speech information in the ASR stage and NLU stage suffering from ASR errors. To solve these problems, end-to-end SLU approach maps speech directly to intent through a single deep learning model. While existing end-to-end SLU models are primarily used for English environment due to lacking large scale SLU dataset in Chinese, and these approaches use only one ASR model to extract features from speech. With the help of Kuaishou technology, a large scale SLU dataset in Chinese is collected to detect abnormal event in their live audio stream. Based on this dataset, this paper proposed a ensemble end-to-end SLU model used for Chinese environment. This ensemble SLU models extracted hierarchies features using multiple pre-trained ASR models, leading to better representation of speech information. This proposed approached achieve 9.7% increase of accuracy compared to previous end-to-end SLU model.

[24] Fei Tao and Carlos Busso. 2019. End-to-end audiovisual speech activity detection with bimodal recurrent neural models. Speech Communication 113 (2019), 25–35.

[25] Fei Tao and Carlos Busso. 2020. End-to-End Audiovisual Speech Recognition System with Multitask Learning. IEEE Transactions on Multimedia (2020).

[26] Seiya Tokui, Ryosuke Okuta, Takuya Akiba, Yusuke Niitani, Toru Ogawa, Shunta Saito, Shuji Suzuki, Kota Uenishi, Brian Vogel, and Hiroyuki Yamazaki Vincent. 2019. Chainer: A Deep Learning Framework for Accelerating the Research Cycle. In Proceedings of the 25th ACM SIGKDD International Conference on Knowledge Discovery & Data Mining. ACM, 2002–2011.

[27] Natalia Tomashenko, Christian Raymond, Antoine Caubrière, Renato De Mori, and Yannick Estève. 2020. Dialogue history integration into end-to-end signal-to-concept spoken language understanding systems. In ICASSP 2020-2020 IEEE International Conference on Acoustics, Speech and Signal Processing (ICASSP). IEEE, 8509–8513.

[28] Pengwei Wang, Liangchen Wei, Yong Cao, Jinghui Xie, and Zaiqing Nie. 2020. Large-Scale Unsupervised Pre-Training for End-to-End Spoken Language Understanding. In ICASSP 2020-2020 IEEE International Conference on Acoustics, Speech and Signal Processing (ICASSP). IEEE, 7999–8003.

[29] Shinji Watanabe, Takaaki Hori, Shigeki Karita, Tomoki Hayashi, Jiro Nishitoba, Yuya Unno, Nelson Enrique Yalta Soplin, Jahn Heymann, Matthew Wiesner, Nanxin Chen, Adithya Renduchintala, and Tsubasa Ochiai. 2018. ESPnet: End-to-End Speech Processing Toolkit. In Proceedings of Interspeech. 2207–2211. https://doi.org/10.21437/Interspeech.2018-1456

[30] Shinji Watanabe, Takaaki Hori, Suyoun Kim, John R Hershey, and Tomoki Hayashi. 2017. Hybrid CTC/attention architecture for end-to-end speech recognition. IEEE Journal of Selected Topics in Signal Processing 11, 8 (2017), 1240–1253.

[31] Haoran Wei, Pranav Chopada, and Nasser Kehtarnavaz. 2020. C-MHAD: Continuous Multimodal Human Action Dataset of Simultaneous Video and Inertial Sensing. Sensors 20, 10 (2020), 2905.

[32] Haoran Wei and Nasser Kehtarnavaz. 2018. Determining number of speakers from single microphone speech signals by multi-label convolutional neural network. In IECON 2018-44th Annual Conference of the IEEE Industrial Electronics Society. IEEE, 2706–2710.

[33] Zhen Zhang, Hao Huang, and Kai Wang. 2020. Using Deep Time Delay Neural Network for Slot Filling in Spoken Language Understanding. Symmetry 12, 6 (2020), 993.

[34] Xinyuan Zhou, Emre Yılmaz, Yanhua Long, Yijie Li, and Haizhou Li. 2020. Multi-Encoder-Decoder Transformer for Code-Switching Speech Recognition. arXiv preprint arXiv:2006.10414 (2020).